\documentclass[12pt]{article}

\newcommand{\Z}{\ensuremath{{\mathsf{Z\!\!Z}}}}

\newcommand{\C}{\ensuremath{{\mathrm{C\hspace{
-1.7mm}\rule{0.3mm}{2.6mm}\;}}}}

\begin{document}

\title{{Dynamical Delocalization for the 1D Bernoulli Discrete  Dirac
Operator}}

\author{C\'esar R de Oliveira and Roberto A Prado\\
\vspace{-0.6cm}\small
\it Departamento de Matem\'{a}tica, UFSCar,
\small \it S\~{a}o Carlos, SP, 13560-970 Brazil\\
\small}
\date{ \today}
\maketitle

\begin{abstract} An {\sc 1D} tight-binding version of  the Dirac
equation is considered; after checking that
it recovers the usual discrete Schr\"odinger equation in the
nonrelativistic limit, it is found that for
two-valued  Bernoulli potentials the zero mass case presents absence of
   dynamical localization for
specific values of the  energy, albeit it has no continuous spectrum.
For
the  other energy  values (again
excluding some very specific ones) the Bernoulli Dirac system is
localized,
independently of the mass.

\end{abstract}

\clearpage

In one-dimensional quantum systems general  random potentials induce
localization and no
conductance~\cite{And,CKM},  irrespective of the disorder intensity.
Exceptions are  restricted to random
models with local correlations, as  polymer models~\cite{JSS}, random
palindrome  models~\cite{CdeO}, both
including the important precursory random  dimers~\cite{F,DPW}
(see~\cite{deBG,JSS} for  rigorous
approaches). In this article it is presented a random discrete model with no
local  correlation for which
delocalization occurs in some  situations; to the best of authors'
knowledge it is the first  tight-binding
model with such property (in \cite{GKT} dynamical delocalization is shown for a tight-binding
Schr\"odinger model, but the random potential is decaying). Since mathematical proofs will appear
elsewhere, it is hoped that the present note will contact physicists with recent interesting mathematical
results on delocalization in one-dimension.

The model is in fact very simple. It is a relativistic  version of the
well-known tight-binding
Schr\"odinger Hamiltonian (with $\hbar=1$)
\begin{equation}
\label{Soperator}  (H_S\psi)_n=-\frac{1}{2m}(\Delta\psi)_n +V_n\psi_n
= \frac{1}{2m}(-\psi_{n+1}
-\psi_{n-1} + 2\psi_n) + V_n\psi_n
\end{equation} (recall that it is common to take $+\Delta$ instead
of~$-\Delta$, and also exclude the constant factor ``2'', in the kinetic term). For
very general random potentials
$V_n$ the model~(\ref{Soperator}) is localized,  including the
Bernoulli potential for which the site
energy $V_n$  is assigned one of two values
$\pm v$ at random ($-v$ with probability $0<p<1$  and~$+v$ with
probability~$1-p$, say); the  spectrum of
the corresponding operator has no continuous  component~\cite{CKM}.

By imposing the strong local correlation that the site  energies $V_n$
are assigned for pairs of lattices,
i.e,
$V_{2n}=V_{2n+1}=\pm v$ for all~$n$, one gets the  random dimer model
exhibiting delocalized
states~\cite{F,DPW,deBG,JSS}. Despite of dynamical  delocalization the
dimer Schr\"odinger operator has  no
continuous component in its spectrum. The first example of system with
dynamical
delocalization and pure point spectrum was a
peculiar almost periodic operator~\cite{delRJLS}. Such  results
clarified the difference between
mathematical localization (i.e., pure point spectrum) and dynamical
localization (i.e., bounded moments,
see ahead). In  the zero-mass case, the Dirac~\cite{BD} model discussed
here has pure  point spectrum and
dynamical delocalization with  no added correlation to the Bernoulli
potential. A crucial first step for
the  arguments will be the appropriate way of writing the transfer
matrices and their similarity with those
of  the Schr\"odinger dimer model.

Previous works have considered the one-dimensional Dirac equation and
relativistic effects on conduction in disordered systems~\cite{RB},
localization~\cite{BRMAMS} and comparative studies
of relativistic and nonrelativistic Kronig-Penney models with
$\delta$-function potentials~\cite{BRMAMS,DAMAR}.
Although tight-binding equations for the electronic amplitudes have
naturally arisen in some of these studies, the
phenomenon
 reported here have not. The interesting question of the comparison of the
relativistic and
nonrelativistic localization length was numerically
investigated~\cite{BRMAMS} in some cases, and it was found that
which one is larger depends on the on-site energy and also on the energy
particle.

Consider a particle of mass~$m\ge0$ in the  one-dimensional
lattice~$\Z$ under the site  potential
$V=(V_n)$; the Dirac tight-binding version is proposed as
\begin{equation}\label{Doperator}  H_D(m,c)=H_0(m,c)+VI_2= \left(
{\matrix{0&c{d}^*\cr  c{d}&0\cr }}
\right) + mc^2\:\sigma_3 +VI_2\, ,
\end{equation} with $c>0$ being the speed of light,
$\sigma_3$ the usual Pauli matrix, $I_2$ the $2\times2$  identity
matrix and
${d}$ a finite difference operator (a discrete  analogue of the first
derivative)
\[ ({d}\psi)_n = \psi_{n+1} - \psi_{n}.
\] Since ${d}$ is not Hermitian, its adjoint
$({d}^*\psi)_n =
\psi_{n-1}-\psi_n$ appears in the definition of
$H_D$ (the inclusion of the imaginary unit $i$ in  front of the
difference operators~$d$ and~$d^*$ is
immaterial).  In case
$V_n$ takes a finite number of values, it is clear that
$H_D$ is a bounded Hermitian operator acting on
$\ell^2(\Z;\C^2)$ and the resulting Dirac equation  can be recast in
the compact form
\begin{equation}\label{Dequation} i\frac{\partial
\Psi_n}{\partial t}=\left(H_D(m,c)\Psi\right)_n= \left(
{\matrix{mc^2+V_n&c{d}^*\cr c{d}&-mc^2+V_n\cr }}
\right)\Psi_n,
\end{equation} with the spinor $\Psi=(\Psi_n)$ and
$\Psi_n=\left( {\matrix{\psi^+_n\cr\psi^-_n\cr}}
\right)$.

One can easily verify that the nonrelativistic limit
of~(\ref{Dequation}) is the equation associated to
the Schr\"odinger operator~(\ref{Soperator}); this is an  important
support for the Dirac model just
introduced. Following the traditional prescription for the
nonrelativistic limit  of the Dirac
equation~\cite{BD}, first one removes the rest energy by inserting
$\Psi=e^{-imc^2t}\Phi=e^{-imc^2t}\left({\matrix{\phi^+\cr
\phi^-}}\right)$ into~(\ref{Dequation})  so that
\[ i\frac{\partial \Phi}{\partial t}= c\left(  {\matrix{{d}^*\phi^-\cr
{d}\,\phi^+}} \right) - 2mc^2
\left({\matrix{0\cr \phi^-}}\right) + V\Phi.
\] For large values of~$c$, the equation in the  second row above can
be solved approximately as
$\phi^-={d}\phi^{+}/2mc$, and inserting this into  the first equation
results in
\begin{equation}\label{NRL+} i\frac{\partial
\phi^+}{\partial t} = \frac{1}{2m} {d}^*{d}\phi^+ +  V\phi^+.
\end{equation} Similarly, by considering
$\Psi=e^{imc^2t}\Phi$ one finds
\begin{equation}\label{NRL-} i\frac{\partial
\phi^-}{\partial t} = -\frac{1}{2m} {d}{d}^*\phi^- +  V\phi^-.
\end{equation} Since ${d}^*{d}={d}{d}^*=-\Delta$,  then~(\ref{NRL+})
and~(\ref{NRL-}) correspond to  the
one-dimensional tight-binding Schr\"odinger  equation associated
to~(\ref{Soperator}) with  positive and
negative free energies, respectively.

Another point directly related to the continuous Dirac  equation is the
presence of the so-called
zitterbewegung~\cite{BD} phenomenon  for~(\ref{Dequation}); here it
will be explicitly considered the
particular case of free particle and  small mass~$m$. Following
Section~69 of Dirac's book~\cite{Dirac}, let
${\hat n}$ denote the position operator $({\hat n}\Psi)_n=n\Psi_n$, so
that its time  evolution under the
free operator $H_0(m,c)$ is
${\hat n}(t)=e^{iH_0 t}{\hat n}e^{-iH_0 t}$; the velocity operator is
then
\[
\frac{d{\hat n}(t)}{dt} = i[H_0,{\hat n}(t)]=e^{iH_0 t}cA(0)e^{-iH_0
t}=cA(t),
\] with $A=A(0)=i\left( {\matrix{0&-{d}^*-1\cr{d}+1&0\cr}}
\right)$. Notice that $A$ is Hermitian, $A^2=I_2$,  so that its
spectrum is
$\pm1$ and then the spectrum of $cA$ is~$\pm c$.  Since $e^{-iH_0 t}$
is unitary, it follows that the
spectrum of the above velocity operator is~$\pm c$ for all~$t$.  Hence
it indicates that the possible speed
measurements would result only in~$\pm c$. Now, for small mass~$m$  the
time derivative of the velocity
operator is given by
\begin{equation}\label{Aoperator}
\frac{d(cA(t))}{dt} = i[H_0,cA(t)]=2iH_0F(t),
\end{equation} with
\[ F(t) = \frac{ic^2}{2}H_0^{-1}e^{iH_0t}\left(
{\matrix{d\:d^*&0\cr0&-{d}^*d\cr}}
\right)e^{-iH_0t}.
\] The operator $F=F(0)$ anticommutes with~$H_0$;  thus
$dF(t)/dt = 2iH_0F(t) $ and it is found that
$F(t)=e^{2iH_0t}F$, which is fast oscillating.  Inserting this
into~(\ref{Aoperator}) one finds
$d(cA(t))/dt=dF(t)/dt$; after integrating from~$0$  to~$t$ one gets
$d{\hat n}(t)/dt=cA-F+e^{2iH_0t}F$ and the velocity  quickly oscillates
around
an average value; this is
a  version  of zitterbewegung.

Now the localization results will be discussed.  Consider
model~(\ref{Doperator}) with $V_n$ taking  the
values~$\pm v$,
$v>0$ randomly. Denote by
$\delta_n^{\pm}$ the elements of the canonical  position basis of
$\ell^2(\Z;\C^2)$, for which all entries are
$\left(\matrix{0\cr 0}\right)$ except at the $n$th entry,  which is
given by $\left(\matrix{1\cr 0}\right)$
and
$\left(\matrix{0\cr 1}\right)$ for the superscript  indices~$+$
and~$-$, respectively. If $\Psi_n=\left(
{\matrix{\psi^+_n\cr\psi^-_n\cr}} \right)$ is a solution  of the
eigenvalue equation
\[ (H_D(m,c)-E)\Psi=0,
\] then it is  simple to check that
\[
\left( {\matrix{\psi^+_{n+1}\cr\psi^-_n\cr}} \right) =  T^E_{V_n} \left(
{\matrix{\psi^+_{n}\cr\psi^-_{n-1}\cr}}
\right),\quad \mbox{with}\quad T^E_{V_n}=\left(
{\matrix{1+\frac{m^2c^4-(E-V_n)^2}{c^2}&\frac{mc^2 +E-
V_n}{c}\cr\frac{mc^2-E+V_n}{c}&1\cr}} \right).
\]
$T^E_{V_n}$ is the transfer matrix at the~$n$th step. Recall  that the
Lyapunov exponent
$\gamma(E)$ represents the average rate of  exponential growing of the
norm of transfer  matrices
\[
\|T^E_{V_n}\cdots T^E_{V_2} T^E_{V_1}\|\approx  e^{\gamma(E)\:n};
\]
$1/\gamma(E)$ is called the localization length. A  vanishing Lyapunov
exponent is an indication of
delocalization, so the next task is to find possible energies~$\tilde
E$  with~$\gamma(\tilde E)=0$. In
order to get  vanishing Lyapunov exponents and diffusion, the arguments
will follow  those
in~\cite{JSS,deBG}; detailed mathematical  proofs will appear
elsewhere~\cite{deOP}.

Given an initial spinor $\Psi$ with only one  nonzero
component (i.e., well-localized in space),
the dynamical delocalization will be probed by the time average of the mean squared
displacement (also called second
dynamical moment)
\[ M^m_\Psi(t) = \frac{1}{t} \int_0^t \sum_n n^2
\left(|\langle\delta^+_n,e^{-iH_D(m,c)s}\Psi
\rangle|^2 + |\langle\delta^-_n,e^{-iH_D(m,c)s}\Psi
\rangle|^2\right) ds;
\] dynamical localization is characterized by a  bounded
$M^m_\Psi(t)\le \mbox{cte},$ for all~$t$; otherwise the system is said to present dynamical delocalization.

First the spectral questions will be faced. By  adapting the multiscale
analysis~\cite{CKM,deBG1}  to the
Dirac operator~(\ref{Doperator}) it is possible to show that, due to the
random character of the Bernoulli potential, for typical realizations
the  spectrum of $H_D(m,c)$ has no continuous component for  any~$m\ge
0$, and with exponentially localized eigenfunctions. In other words, mathematical
localization holds for~$H_D$.

By using as the main tool Furstenberg  theorem~\cite{Furs}, for $m>0$
it is shown that
$\gamma(\tilde E)=0$ if, and only if,
$\tilde E=0$ (for $v=c\sqrt{2+m^2c^2}$) and the four possibilities
$\tilde E = \pm c/\sqrt2\pm
c\sqrt{2+m^2c^2}$ (for
$v=c/\sqrt 2$); so,  for other values of energies dynamical
localization  can be shown. For such
$\tilde E$ values with $\gamma(\tilde E)=0$, it was  not possible to
give an answer about dynamical
localization yet.

Nevertheless, restricted to the massless ($m=0$) case, if $0<v\le c$
the Lyapunov exponent vanishes for
$\tilde E=\pm  v$ and $v\ne c/\sqrt2$, and following~\cite{JSS} it is
possible to  show that 
\[
M^0_\Psi(t)\ge\mbox{cte}\, t^{3/2},
\]
 i.e., there is no dynamical localization despite
the absence of a  continuous
component in the spectrum of the  random operator~$H_D(0,c)$. Due to
its importance here, it is
worth including the main argument for  the vanishing of~$\gamma(E=v)$
(the case
$E=-v$ being similar). In this case, the possible  transfer matrices are
\[ T^v_{-v} =
\left({\matrix{1-\left(\frac{2v}{c}\right)^2&\frac{2v}{c}\cr-
\frac{2v}{c}&1\cr}}\right),\quad \quad T^v_v = I_2.
\] Notice that $T^v_{-v}$ and $T^v_{v}$ are  commuting matrices and
both have spectral radius equal to~1 (for such $v$). If
$n_{-}$ denotes the average number of times that the potential
$-v$ occurs in
$n$ trials, then
$n-n_{-}$ is the average number of times that the  transfer matrix is
the identity. Thus, if $p$ is the
probability for the potential value~$-v$,
\[
\gamma(v)=\lim_{n\to\infty} \frac{1}{n} \ln
\|T^E_{V_n}\cdots T^E_{V_2}  T^E_{V_1}\|=\lim_{n\to\infty}
\frac{n_{-}}{n}\ln\|(T^v_{-v})^{n_{-}}\|^{1/n_{-}}= p\ln  1=0.
\]
The heuristics for the dynamical delocalization in this case can be
found in the paper by Dunlap,
Wu and Phillips in Ref.~\cite{DPW}. The main concern for the proof is the
uniform boundedness of the product of
transfer matrices~\cite{JSS,deOP}. It is very important to stress that here delocalization is not synonymous
of zero Lyapunov exponent, as some people have considered.

For small but nonzero mass, it is expected that the  dynamics follow
closely the massless case, at  least
for a small period of time. The final result to be reported is an
inequality confirming such expectative;
by making  using of Duhamel formula, it can be shown that, given an initial~$\Psi$, there exists
$C>0$ so that, for all  $t>0$,
\begin{equation}\label{zeroNzero}
\left|M^0_\Psi(t)-M^m_\Psi(t) \right|\le C\,mc^2t^4.
\end{equation} Therefore, if the time $t$ is not too  large and/or the
mass $m$ is sufficiently small, the
mean squared displacement follows rather closely the  delocalized
massless case, so that inattentive
numerical simulations could give a wrong insight.

It is natural that this model would be applied to any case the one-dimensional tight-binding Schr\"odinger
operator was used; it would be the first step for their relativistic versions. Since the Dirac operator
in the massless case presents dynamical delocalization, this becomes a potential source for explaining some
observed effects (at least for small~$m$) as details in the theory of mesoscopic systems~\cite{I}.

Summing up, a natural one-dimensional Dirac tight-binding model  was
proposed which was
supported by its  nonrelativistic
limit (it recovers the discrete Schr\"odinger model) and the  presence
of zitterbewegung. Then results
about  mathematical and dynamical localization were reported for such
operator with random Bernoulli
potentials: for all  values of $c>0$ and mass $m\ge0$, there is
mathematical localization,  but in the
massless case and potential intensity
$v\le c$, particular values of the energy imply the absence of
dynamical localization, although no
potential correlation was imposed. It is possible that this  model is
the simplest one with such
delocalization.  Finally,  relation~(\ref{zeroNzero}) gives
quantitatively an  estimate of how, for small
time~$t$, the dynamics of  the localized regime follows the delocalized
one.

\subsubsection*{Acknowledgments} {\small  CRdeO thanks the partial
support by CNPq, RAP  was supported by
CAPES} (Brazilian government agencies).

\end{document}